\documentclass{aastex63}

\usepackage[version=3]{mhchem}
\usepackage{booktabs}
\usepackage{color,soul}
\usepackage{savesym}
\savesymbol{tablenum}
\usepackage{siunitx}
\restoresymbol{SIX}{tablenum}
\DeclareSIUnit{\molecule}{molecule}
\definecolor{mypink}{rgb}{0.858, 0.188, 0.478}
\definecolor{mypink3}{cmyk}{0, 0.8808, 0.9429, 0.3412}
\definecolor{mygreen}{cmyk}{0.59, 0.00, 0.99, 0.10}
\definecolor{lightblue}{rgb}{0.659,0.8706,1}
\definecolor{blu}{rgb}{0.2039,0.388,1}
\sethlcolor{lightblue} \setstcolor{red}

\renewcommand{\thefootnote}{\fnsymbol{footnote}}
\renewcommand\arraystretch{1.2}
\submitjournal{ApJ}

\shorttitle{\ce{HCOCN} gas phase formation route}
\shortauthors{Tonolo et al.}
\graphicspath{{./}{figures/}}

\begin{document}

\title{The quest for a plausible formation route of formyl cyanide in the interstellar medium: a state-of-the-art quantum-chemical and kinetic approach}

\correspondingauthor{Vincenzo Barone}
\email{vincenzo.barone@sns.it}
\correspondingauthor{Cristina Puzzarini}
\email{cristina.puzzarini@unibo.it}

\author[0000-0002-9555-7834]{Francesca Tonolo}
\affiliation{Scuola Normale Superiore,
Piazza dei Cavalieri 7,
Pisa, 56126, Italy}
\affiliation{Department of Chemistry ``Giacomo Ciamician'', University of Bologna,
Via F. Selmi 2,
Bologna, 40126, Italy}

\author[0000-0001-6522-9947]{Jacopo Lupi}
\affiliation{Scuola Normale Superiore,
Piazza dei Cavalieri 7,
Pisa, 56126, Italy}

\author[0000-0002-2395-8532]{Cristina Puzzarini}
\affiliation{Department of Chemistry ``Giacomo Ciamician'', University of Bologna,
Via F. Selmi 2,
Bologna, 40126, Italy}

\author[0000-0001-6420-4107]{Vincenzo Barone}
\affiliation{Scuola Normale Superiore,
Piazza dei Cavalieri 7,
Pisa, 56126, Italy}



\begin{abstract}
Interstellar complex organic molecules (iCOMs) are assumed to be mainly formed on dust-grain surfaces. However, neutral gas-phase reactions in the interstellar medium (ISM) can play an important role. In this paper, by investigating the reaction between aldehydes and the cyano radical, we show that both formaldehyde (\ce{CH2O}) and acetaldehyde (\ce{CH3CHO}) can lead to the formation of formyl cyanide (HCOCN). Owing to accurate quantum-chemical computations followed by rate constant evaluations, we have been able to suggest and validate an effective mechanism for the formation of HCOCN, one of the molecules observed in the ISM. Quite interestingly, the mechanism starting from \ce{CH2O} is very effective at low temperature, while that involving \ce{CH3CHO} becomes more efficient at temperatures above 200 K.
\\
\end{abstract}

\keywords{ISM: abundances – ISM: molecules}


\section{Introduction} \label{sec:intro}
In the last 50 years more than 200 molecular species have been identified in the interstellar medium (ISM) mostly thanks to their rotational signatures \citep{McGuire_2018}. Among them, the so-called interstellar complex organic molecules (iCOMs), namely molecules containing at least one carbon atom and --overall-- more than 5 atoms, are particularly significant because they include several precursors of biomolecule building blocks \citep{exo,COMs,horst12,C2CS35113G,Saladino-RSC2012_Formamide,SaladinoE2746}. 

The most widespread investigated reactions to form iCOMs are based on grain-surface chemistry because radical species trapped in icy-mantles can easily react and give rise to a rich chemistry (see, e.g., \cite{Garrod06,Garrod_2008,Oberg_2010,0144235X.2015.1046679,acsearthspacechem.7b00156}). However, the recent observation of iCOMs also in very cold objects, has suggested that the role gas-phase reactions could have been overlooked (see, e.g., \cite{Bacmann12,Vasyunin_2013,Vastel_2014,slv009}). The extreme conditions of the ISM (i.e. very low temperatures and very low density) pose strong limitations on the feasibility of reactivity in the gas phase, thus requiring that all transition states are submerged with respect to reactants' energy. This gives rise to a chemistry sensibly different from the conventional reactivity, under terrestrial conditions, we were used to. As a consequence, the formation pathways of the detected molecules in the typically cold and (largely) collision free environment of the ISM are often unknown. Aiming at their disclosure and understanding, new reaction mechanisms need to be investigated from both thermochemical and kinetic points of view. While energetic studies allow for deriving possible accessible pathways, only the determination of accurate reaction rates can confirm which formation routes can indeed occur. These in turn provide useful information in view of rationalizing the molecular abundances observed in interstellar clouds. 

Among the most plausible gas-phase mechanisms, radical-neutral reactions play a central role and, among them, the addition of the cyano radical (CN) to electrophilic sites is gaining an increasing interest because it might lead to the formation of compounds containing the --CN-- moiety. The interest on the latter species is due to the fact that they can be key intermediates in the synthesis of aminoacids, the building blocks of proteins, and/or nucleobases, the essential components of nucleic acids \citep{Vastel_2014,slv009,D0CP00561D,acsearthspacechem.0c00062}.  

Despite the fact that the required spectroscopic parameters of formylcyanide (HCOCN) were available since 1995 \citep{BOGEY1995344}, its first and only detection in the ISM is quite recent \citep{Remijan_2008}. In 2008, \cite{Remijan_2008} indeed detected some of its rotational transitions toward the star-forming region Sagittarius B2(N) using the 100 m Green Bank Telescope (GBT). 
In spite of several spectroscopic studies carried out on HCOCN and some studies on its formation in the ISM (see \cite{2013MNRAS.433.3152D} and references therein), to the best of our knowledge, an accurate investigation of possible gas-phase formation pathways 
has not been carried out yet. This prompted us to try to fill this lack. 

The choice of possible precursors for the derivation of a gas-phase formation pathway for formyl cyanide was guided by assumptions related to the physical conditions and molecular abundances of Sagittarius B2(N). Among the various possibilities, the attack of the CN radical to formaldehyde seems to be particularly promising. Indeed, both these molecular species are widely spread in the ISM and their abundance in the Sagittarius B2 nebula qualifies them as excellent candidates \citep{mehringer19956, savage2002galactic}. Furthermore, we extended such an investigation by replacing formaldehyde with acetaldehyde in the reaction with CN, whose widespread presence in the ISM, and --in particular-- in the Sagittarius B2 nebula \citep{Matthews1985}, makes this second approach appealing as well. 

The manuscript is organized as follows. First, the essential computational details for both the energetic and kinetic study are provided. In the subsequent section, the results are reported and analyzed in detail. The outcomes for the investigation of the reaction of formaldehyde and acetaldehyde with the cyano radical are first presented from a thermochemical point of view. Then, kinetic results for both reactions are discussed. Finally, concluding remarks are provided.

\section{Computational methodology} 
\label{sec:compdet}

The starting point for the study of the formation of HCOCN is the identification of the potential reactants and the analysis of the corresponding reactive potential energy surface (PES), which implies the characterization of all stationary points from both a structural and energetic point of view. The accurate thermochemical characterization requires then to be followed by kinetic calculations. In the following the essential details are provided, while a deeper account is given in the Appendix.

As mentioned above, the conditions of the ISM are extreme with regard to physical conditions: low temperatures (10-100 K) and low density (10-10$^7$ particles cm$^{-3}$). By translating density in terms of pressure, a density of 10$^4$ particles cm$^{-3}$ corresponds to a pressure of 3.8$\times$10$^{-10}$ Pa ($\sim$3.8$\times$10$^{-15}$ atm). Therefore, in the investigation of the reactions between formaldehyde or acetaldehyde with CN we take such constraints into considerations.

\subsection{Reactive potential energy surface}
\label{pes:det}

This type of study requires the application of different levels of theory that combine accuracy and efficiency. Indeed, a preliminary investigation of the reactive PES is carried out at an affordable computational cost, thus leading to the evaluation of all possible reaction paths. Then, a structural and energetic characterization of the most favoured paths is performed at a higher level of theory. 

The approach followed is that employed in \cite{Vazart2016} for the gas-phase formation route of formamide and consists of the following steps: \\
(i) The reactive PES has been first of all studied using a cost effective level of theory in order to locate the stationary points. The hybrid B3LYP functional \citep{{Becke1993},{Lee1988}} in conjunction with a double-$\zeta$ quality basis set (SNSD; \cite{Barone2008}) has been employed, with dispersion effects taken into account (D3BJ; \cite{Grimme2010,Grimme2011}). 
\\
(ii) To obtain a more accurate description of these pathways, all stationary points have been re-investigated at a higher level of theory using the double-hybrid B2PLYP functional \citep{doi:grimme2006} (combined with D3BJ corrections) in conjunction with a triple-$\zeta$ quality basis set incorporating diffuse functions (maug-cc-pVTZ-\emph{d}H; \cite{Fornaro2016,papajak2009}).
\\
(iii) To further improve the energy determination of the stationary points of the most energetically favoured paths, single-point energy calculations, at the B2PLYP/maug-cc-pVTZ-\emph{d}H geometries, have been performed by means of the so-called CCSD(T)/CBS+CV composite scheme \citep{Miriam2,hosop,Barone2013}, which is based on coupled-cluster (CC) theory and is explained in the Appendix. 
Finally, CCSD(T)/CBS+CV energies have been combined with anharmonic zero-point energy (ZPE) corrections evaluated at the B2PLYP/maug-cc-pVTZ-\emph{d}H level, as detailed in the Appendix. 
\\

\subsection{Kinetic models}
\label{Kin:Mod}

For both \ce{CH2O + CN} and \ce{CH3CHO + CN} reactions, global rate constants have been calculated by using a master equation (ME) approach based on \emph{ab initio} transition state theory (AITSTME), thereby employing the MESS software as master equation solver (\cite{georgievskii2013reformulation}). For elementary reactions involving a transition state, rate constants have been computed using transition state theory (TST), while for barrierless elementary reactions, they have been evaluated by means of phase space theory (PST; \cite{pechukas1965detailed,chesnavich1986multiple}). Tunnelling has been accounted for using the Eckart model \citep{eckart1930penetration}. A more detailed account is provided in Appendix \ref{appendix:kin}.

\section{Results and discussion} \label{sec:results}

In Section \ref{4.1}, the results for the reactive PES of \ce{CH2O + CN} are reported and discussed, mainly focusing on the energetically favoured paths. Subsequently, the analogous investigation on the reaction between acetaldehyde and CN is detailed in Section \ref{4.2}. The outcomes of the kinetic study for both reactions are reported and discussed in Section \ref{4.3}.

\subsection{Mechanistic Study: CN + formaldehyde}\label{4.1}

The examined reaction paths include:
(i) the nucleophilic attack by carbon-CN on the carbonyl oxygen;
(ii) the nucleophilic attack of the CN radical (both C-approach and N-approach) on the carbonyl carbon of formaldehyde;
(iii) the abstraction of a hydrogen atom of formaldehyde by the CN radical.

\begin{figure}
\begin{center}
\includegraphics[scale=0.22]{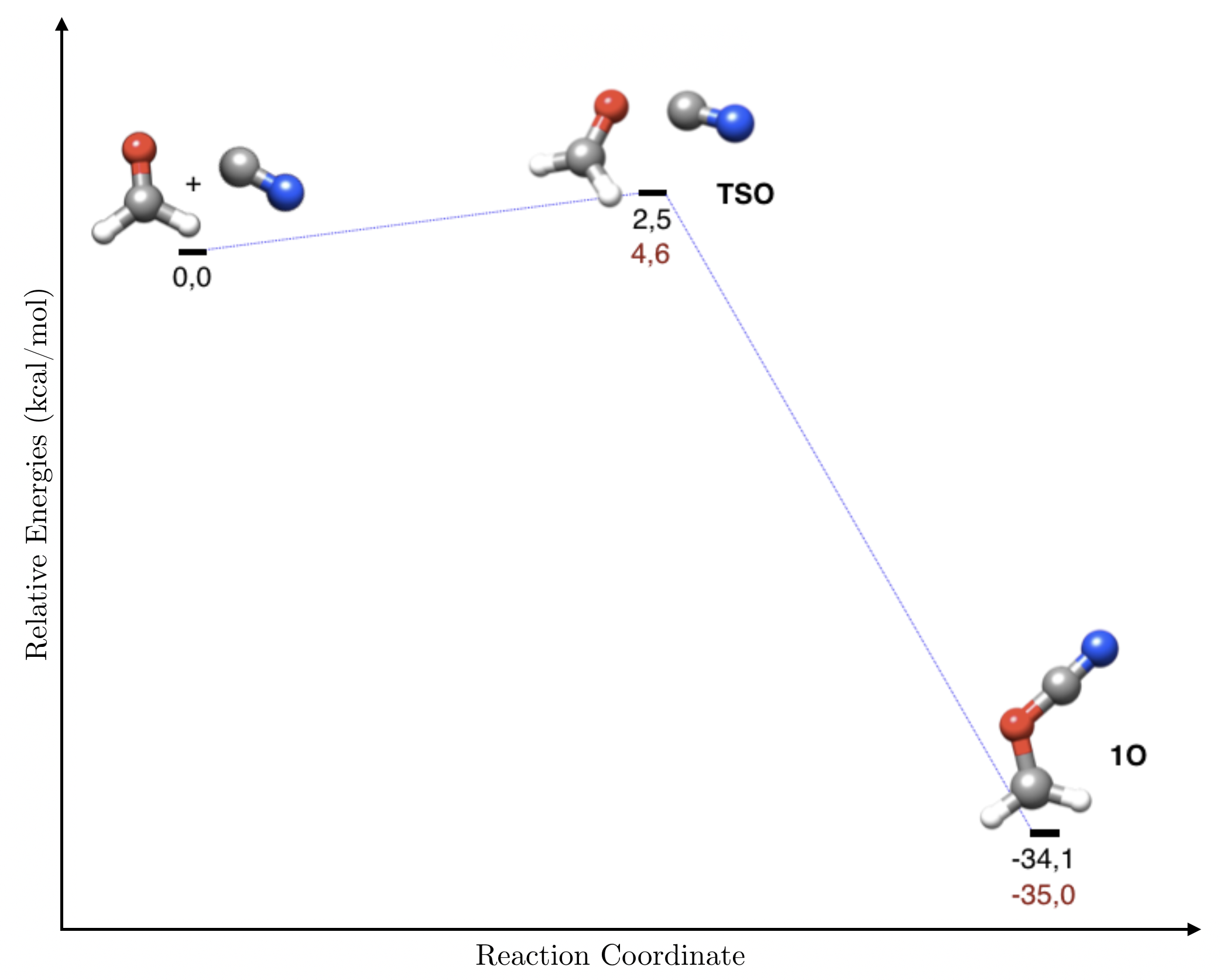}
\end{center}
\caption{Reaction path of \ce{CN + CH2O} for the attack of the CN radical on the oxygen side: relative electronic B2PLYP/maug-cc-pVTZ-\emph{d}H (\textbf{black}) and harmonic ZPE-corrected (\textcolor{mypink3}{dark-red}) energies are reported.}
\label{fig:5}
\end{figure}

As expected, the radical attacks on the carbonyl carbon site are more favorable than those on oxygen because of their different electrophilicity; in addition, the CN radical attacks preferentially by the carbon side where the unpaired electron is more localized.
For these reasons, the N-attack by the CN radical on oxygen of formaldehyde is so unfavorable that it does not even occur.
Likewise, the C-attack by CN on the oxygen atom of formaldehyde would lead to the formation of a single product, shown in Figure \ref{fig:5}, through a transition state at higher energy than the reactants, which cannot be overcome in the typical conditions of the ISM. Therefore, this pathway has not been further considered.
\begin{figure*}
\begin{center}
\includegraphics[scale=0.40]{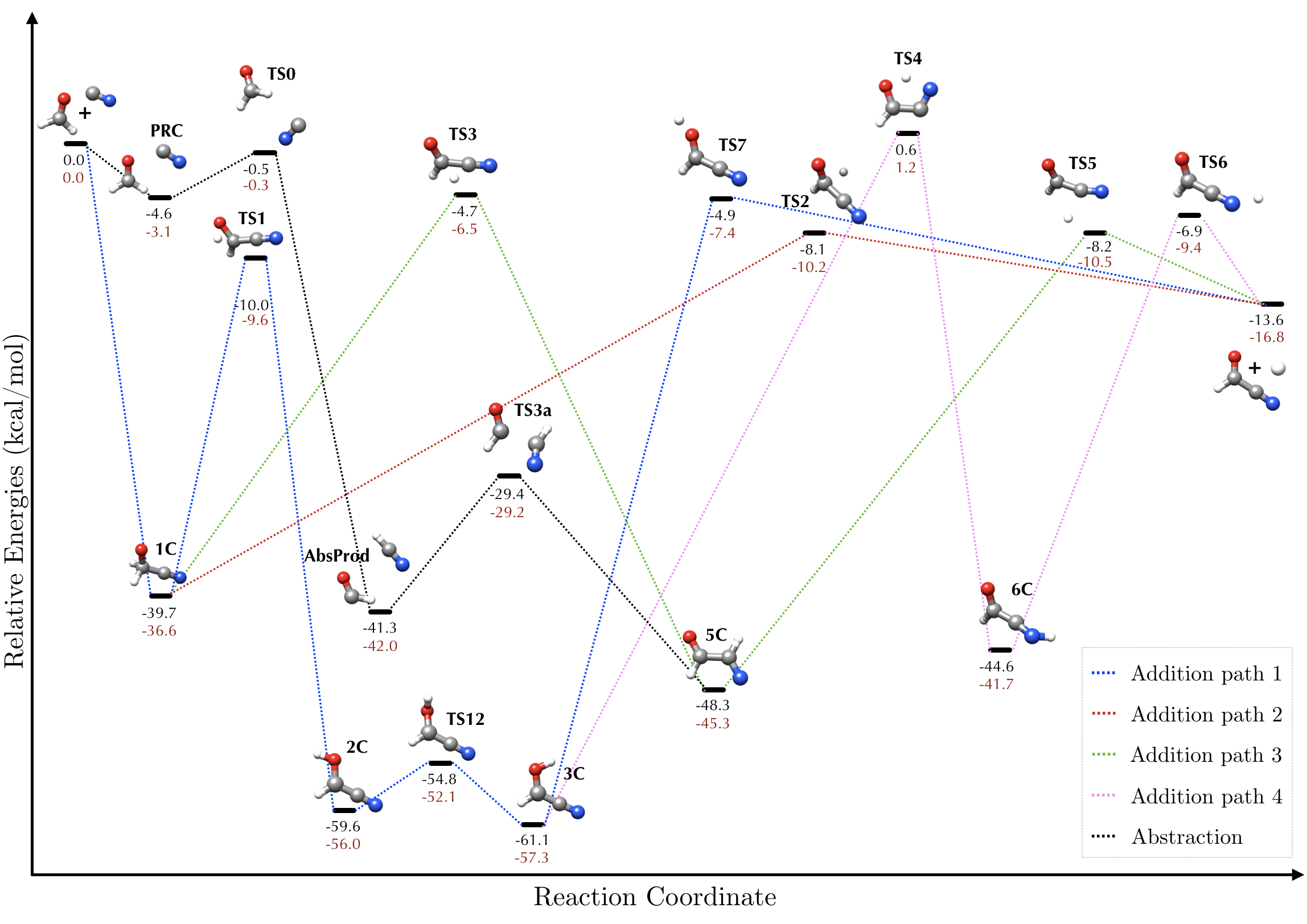}
\caption{Formation route of formyl cyanide from the C-attack of the CN radical on the carbon side of \ce{CH2O}: relative CCSD(T)/CBS+CV (\textbf{black}) single-point energies and anharmonic ZPE-corrected (\textcolor{mypink3}{dark-red}) energies are reported.}
 \label{fig:1}
\end{center}
\end{figure*}

Figure \ref{fig:1}
shows the reaction paths derived for the C-attack of the CN radical on formaldehyde and the abstraction of a hydrogen atom of formaldehyde by the CN radical. 
These pathways are those of interest because they can lead to the formation of the sought product only requiring submerged barriers (i.e. lower in energy with respect to reactants) to be overcome.

The first molecular species formed from the interaction between the two reactants, is the tetrahedral intermediate \textbf{1C}, which is energetically favored by about 40 kcal/mol and can lead to the desired products overcoming the transition state \textbf{TS2} (see red path). This path is particularly interesting from an energetic point of view because it yields formyl cyanide by passing through the least number of intermediates and transition states. The intermediate \textbf{1C} can also lead to the formation of the two isomers \textbf{2C} and \textbf{3C} which, being the most stable intermediates of the investigated reactive PES, could be competitive with the formation of formyl cyanide. The same is for the green path, which is also favored from an energetic point of view, but it might represent a kinetic sink for the sought reaction. The path represented in pink is ruled by a transition state (\textbf{TS4}) lying at an energy very close to that of the reactants. Therefore, this path is not very favourable under the ISM conditions. Finally, the black path involves the abstraction of a hydrogen of formaldehyde by the CN radical: a pre-reactive complex (\textbf{PRC}) is formed, and then the abstraction occurs. After overcoming the transition state \textbf{TS0}, PRC leads to the formation of HCN and the formyl radical (HCO), denoted as \textbf{AbsProd}. The transition state \textbf{TS0} is slightly more stable than the reactants also at higher levels of theory (vide infra).
Once the \textbf{AbsProd} are formed, they can further react to give the intermediate \textbf{5C} $via$ the transition state \textbf{TS3a}, thus leading to the formation of formyl cyanide along the green path. The involvement of these intermediates in further chemical transformations would be probably possible only via a roaming mechanism \citep{B512847C,ar8000734}. However, in view of the large excess of energy and the high barrier for the subsequent transition state relative to the abstraction products (TS3a), the contribution of this reaction channel to the overall reaction is negligible.
\begin{table}
\centering
\scriptsize
\vspace{0.5cm}
 \caption{Relative electronic energies (in kcal/mol) of the stationary points shown in Figure \ref{fig:1}, calculated at the B2PLYP/maug-cc-pVTZ-\emph{d}H and CCSD(T)/CBS+CV levels of theory, also including anharmonic ZPE corrections.}
 \label{tab1}
 \begin{tabular}{ c c c c c }
\toprule
 & \multicolumn{2}{c}{B2PLYP/maug-cc-pVTZ-\emph{d}H} & \multicolumn{2}{c}{CCSD(T)/CBS+CV} \\ 
\cline{2-5} 
 & Energy     & ZPE Corrected   & Energy     & ZPE Corrected \\
 & (kcal/mol) &  (kcal/mol)     &(kcal/mol) &  (kcal/mol)  \\
\midrule
\ce{CH2O} $+$ CN  &  \phantom{-0}0.0 &  \phantom{-0}0.0  &  \phantom{-0}0.0  & \phantom{-0}0.0  \\
1C              &  -41.1 &  -38.0  &    -39.7  & -36.6  \\
PRC           &  \phantom{0}-7.1   &  \phantom{0}-5.4    &    \phantom{0}-4.6     &  \phantom{0}-3.1  \\
TS1             &  -11.5 &  -11.1  &    -10.0  & \phantom{0}-9.6  \\
TS0            &   \phantom{0}-0.8   &  \phantom{0}-0.5   &    \phantom{0}-0.5    &  \phantom{0}-0.3 \\
AbsProd   &  -43.2  &  -43.8  &   -41.3  &   -42.0 \\
2C              &  -60.8 &  -57.2  &    -59.6  & -56.0  \\
TS12            &  -55.7 &  -53.0  &    -54.8  & -52.1  \\
3C              &  -62.4 &  -58.6  &  -61.1  & -57.3  \\
TS3             &  \phantom{0}-7.4  &  \phantom{0}-9.2   &    \phantom{0}-4.7   & \phantom{0}-6.5  \\
TS3a           & -32.3   &  -31.9  &  -29.4    &  -29.2  \\
5C              &  -49.8 &  -46.7  &    -48.3  & -45.3  \\
TS7             &  \phantom{0}-8.3  &  -10.8  &    \phantom{0}-4.9  &  \phantom{0}-7.4  \\
TS2             &  -10.4 &  -12.5  &   \phantom{0}-8.1  &  -10.2  \\
TS4             &  \phantom{0}-1.3  &  \phantom{0}-0.7   &     \phantom{-0}0.6  &  \phantom{-0}1.2  \\
6C              &  -48.2 &  -45.3  &    -44.6  & -41.7  \\
TS5             &  -10.3 &  -12.6  &    \phantom{0}-8.2  &  -10.5  \\
TS6             &  -10.6 &  -13.0  &   \phantom{0}-6.9  &  \phantom{0}-9.4  \\
HCOCN $+$ H     &  -15.3 &  -18.5  &    -13.6  & -16.8  \\
\bottomrule
\end{tabular}
\end{table}

\begin{figure}
\begin{center}
 \includegraphics[scale=0.27]{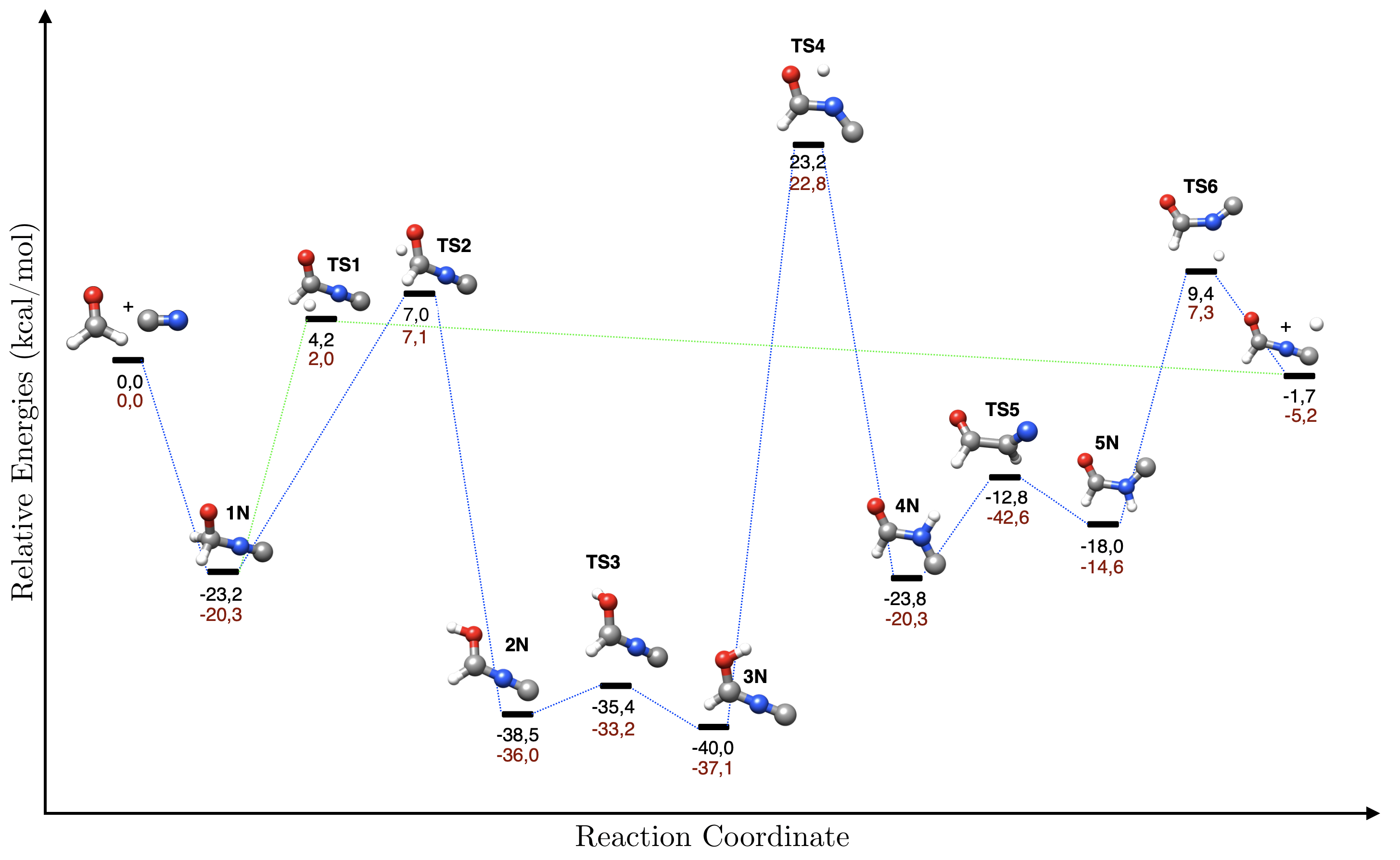}
 \end{center}
 \caption{Reaction path of the reaction \ce{CN + CH2O} $\rightarrow$ HCONC $+$ H for the attack of the CN radical on the nitrogen side: relative electronic B2PLYP/maug-cc-pVTZ-\emph{d}H (\textbf{black}) and harmonic ZPE corrected (\textcolor{mypink3}{dark-red}) energies are reported.}
 \label{fig:2}
\end{figure}

From an energetic point of view, the various pathways of Figure \ref{fig:1} are open even in the harsh conditions of the ISM. For this reason, they have been 
investigated at a higher level of theory (CCSD(T)/CBS+CV), the results being reported in Table \ref{tab1}.
While this composite scheme does not change the trends resulting from B2PLYP calculations, it reduces the relative energies with respect to the reactants. This leads to intermediates that are less stable by about 1-3 kcal/mol, which often means small energy barriers to be overcome.   
Based on the available literature (see, e.g., \cite{hosop,Barone2013,D0CP00561D}), the CCSD(T)/CBS+CV approach improves the expected accuracy, thus reducing the error bars on relative energies below 0.5 kcal/mol.

The last mechanism considered for the reaction between CN and \ce{CH2O} is the N-attack of the cyano radical on the carbon side of formaldehyde, which is shown in Figure \ref{fig:2}. 
All the possible paths issuing from this approach involve transition states at higher energy than the reactants, thus preventing the reaction to occur in the typical conditions of the ISM.

\subsection{Mechanistic Study: CN + acetaldehyde}\label{4.2}

\begin{figure}
\begin{center}
\includegraphics[scale=0.32]{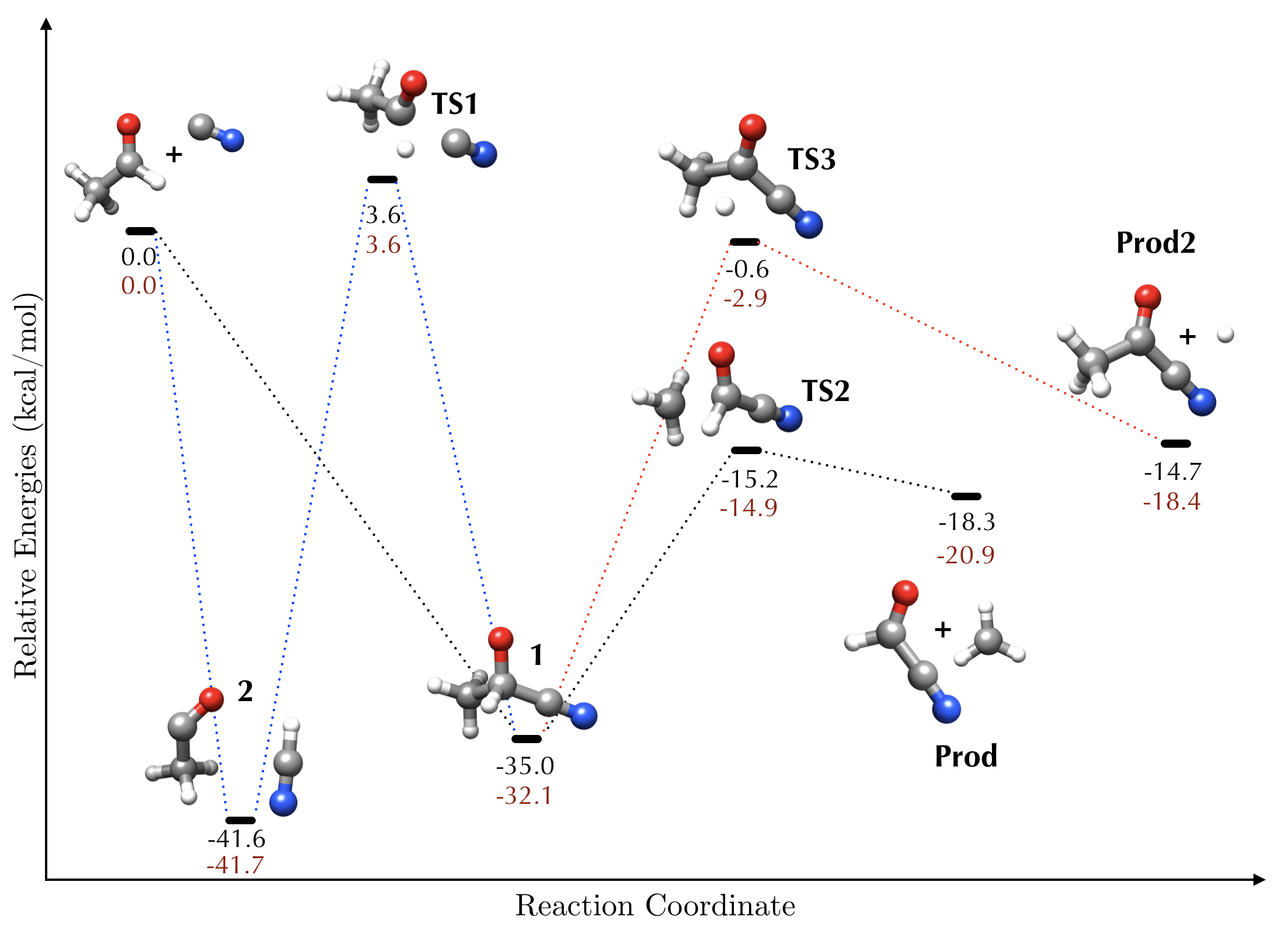}
\caption{Formation route of formyl cyanide from the C-attack of the CN radical on the carbon side of \ce{CH3CHO}: relative CCSD(T)/CBS+CV (\textbf{black}) single-point energies and anharmonic ZPE-corrected (\textcolor{mypink3}{dark-red}) energies are reported.}
 \label{fig:wa}
\end{center}
\end{figure}

As mentioned in the Introduction, another mechanism investigated for the formation of formyl cyanide is the reaction between acetaldehyde and the cyano radical. 
Figure \ref{fig:wa} shows the corresponding reactive PES.

The path identified in black represents the direct attack of CN on the carbonyl carbon of acetaldehyde. In analogy to the \ce{CH2O + CN} reaction, this attack
leads to the formation of a tetrahedral intermediate (\textbf{1}), from which both formyl cyanide and acetyl cyanide (red path) can be obtained. 
Since the formation of both products involves only submerged barriers, the corresponding paths are feasible from an energetic point of view in the conditions of the interstellar medium. Furthermore, it is interesting that formyl cyanide is the favored product from an energetic point of view.  
The path traced in blue describes the abstraction of a hydrogen atom of acetaldehyde by the CN radical. This leads to the formation of HCN and the acetyl radical (\ce{CH3CO}), which can further react and form the tetrahedral intermediate \textbf{1} through the transition state \textbf{TS1}. In analogy to the hydrogen abstraction path of the \ce{CH2O + CN} reaction, roaming mechanisms might play a role \citep{B512847C,ar8000734}. However, as above, the large excess of energy and the high barrier due to TS1 are expected to provide a negligible contribution.

As already discussed for \ce{CH2O + CN}, for all stationary points, improved energies have been obtained by single-point CCSD(T)/CBS+CV computations at B2PLYP/maug-cc-pVTZ-$d$H reference geometries. The corresponding results, collected in Table \ref{tab2}, confirm the overall picture provided by B2PLYP energies. However, it is noteworthy that the energy of \textbf{TS1} at such level of theory is higher than that of the reactants, thus preventing the H abstraction to occur in the typical conditions of the ISM. As a general note, a brief comment on the comparison on B2PLYP and CCSD(T)/CBS+CV energies is deserved. For all the stationary points characterizing the addition of CN to formaldehyde, the maximum and average absolute errors of B2PLYP results with respect to CCSD(T)/CBS+CV values are 3.8 and 2.1 kcal/mol, respectively. Moving to the addition to acetaldehyde, the deviations remain similar, i.e. 3.7 and 1.7 kcal/mol, respectively. Overall, the B2PLYP/maug-cc-pVTZ-dH level leads to errors about four times larger than those expected for the CCSD(T)/CBS+CV approach.

\begin{table}
\centering
\scriptsize
\setlength{\tabcolsep}{0.3pt}
 \caption{Relative electronic energies (in kcal/mol) of the stationary points shown in Figure \ref{fig:wa}, calculated at the B2PLYP/maug-cc-pVTZ-\emph{d}H and CCSD(T)/CBS+CV levels of theory, also including anharmonic ZPE corrections.}
 \label{tab2}
 \begin{tabular}{ c c c c c }
\toprule
 & \multicolumn{2}{c}{B2PLYP/maug-cc-pVTZ-\emph{d}H} & \multicolumn{2}{c}{CCSD(T)/CBS+CV} \\ 
\cline{2-5} 
 & Energy     & ZPE Corrected   & Energy     & ZPE Corrected \\
 & (kcal/mol) &  (kcal/mol)     &(kcal/mol) &  (kcal/mol)  \\
\midrule
CH$_3$COH $+$ CN  &  \phantom{-0}0.0 &  \phantom{-0}0.0  &  \phantom{-0}0.0  & \phantom{-0}0.0  \\
1                 &  -35.7           &  -32.8            &    -35.0          & -32.1            \\
2                 &  -43.5          &   -43.5            &    -41.6          &  -41.7           \\ 
TS1            &  \phantom{0}-0.3             &   \phantom{0}-0.4             &    \phantom{0}3.6             &    \phantom{0}3.6            \\
TS2               &  -18.0           &  -17.8            &    -15.2          & -14.9            \\
HCOCN $+$ CH$_3$  &  -20.4           &  -23.1            &    -18.3          & -20.9            \\
TS3                             & \phantom{0}-8.8     &   -11.1      &    \phantom{0}-0.6       &   \phantom{0}-2.9    \\
CH$_3$COCN $+$ H &  -16.2    &  -19.9       &   -14.7        &   -18.4    \\
\bottomrule
\end{tabular}
\end{table}

\subsection{Rate coefficients}\label{4.3}

According to the results discussed in the two preceding sections, the reactive PES for the formation of formyl cyanide from \ce{CH3CHO} and CN involves intermediates that are --in the majority of the cases-- energetically less stable than those of the \ce{CH2O + CN} PES. Therefore, a kinetic study is mandatory, not only to evaluate the specific (and global) rate constants, but also to understand whether: \\
(i) the presence of particularly stable intermediates in the \ce{CH2O + CN} reaction might represent a kinetic sink; \\
(ii) both reactions are effective and/or which one is more efficient; \\
(iii) concerning point (ii), the temperature plays a role.

\begin{figure}[htbp]
\begin{center}
 \includegraphics[scale=1.5]{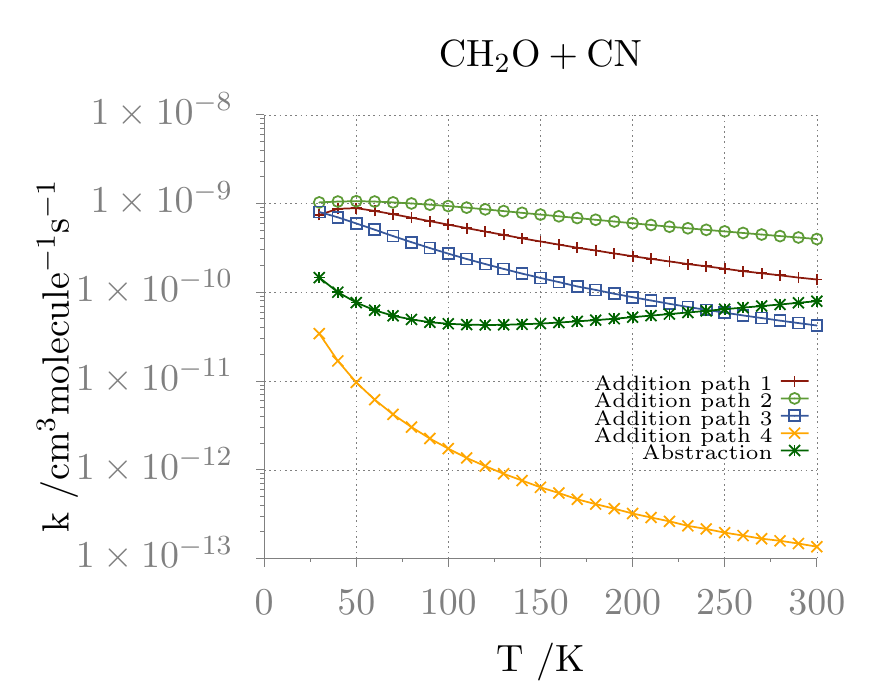}
 \end{center}
 \caption{Channel specific rate constants for the \ce{CH2O + CN} reaction leading to \ce{HCOCN + H}.}
 \label{fig:rateform}
\end{figure}

\begin{figure}
\begin{center}
 \includegraphics[scale=1.5]{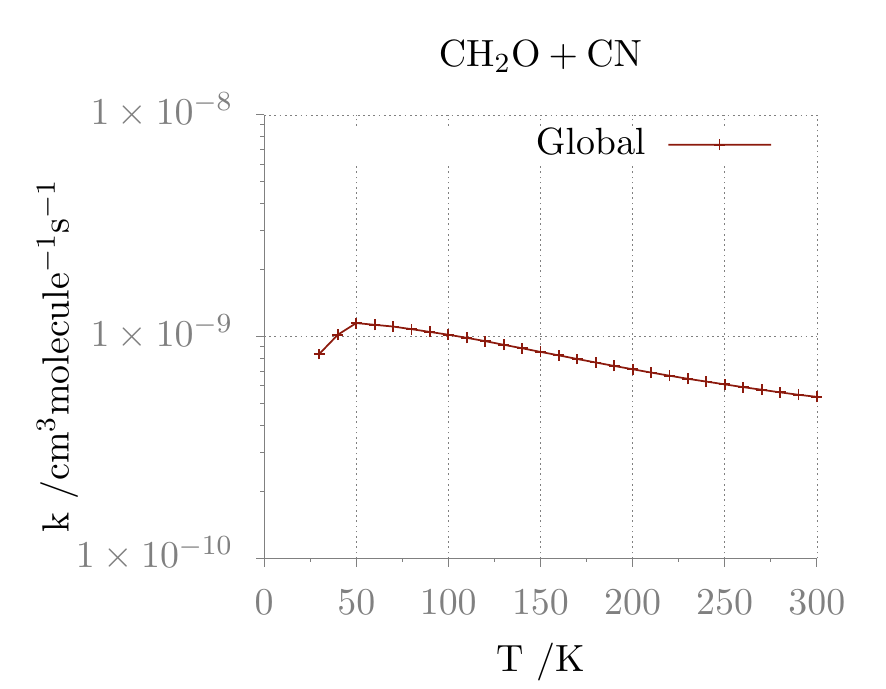}
 \end{center}
 \caption{Global rate constant for \ce{CH2O + CN} reaction.}
 \label{fig:rateformglobal}
\end{figure}

\begin{figure}
\begin{center}
 \includegraphics[scale=1.5]{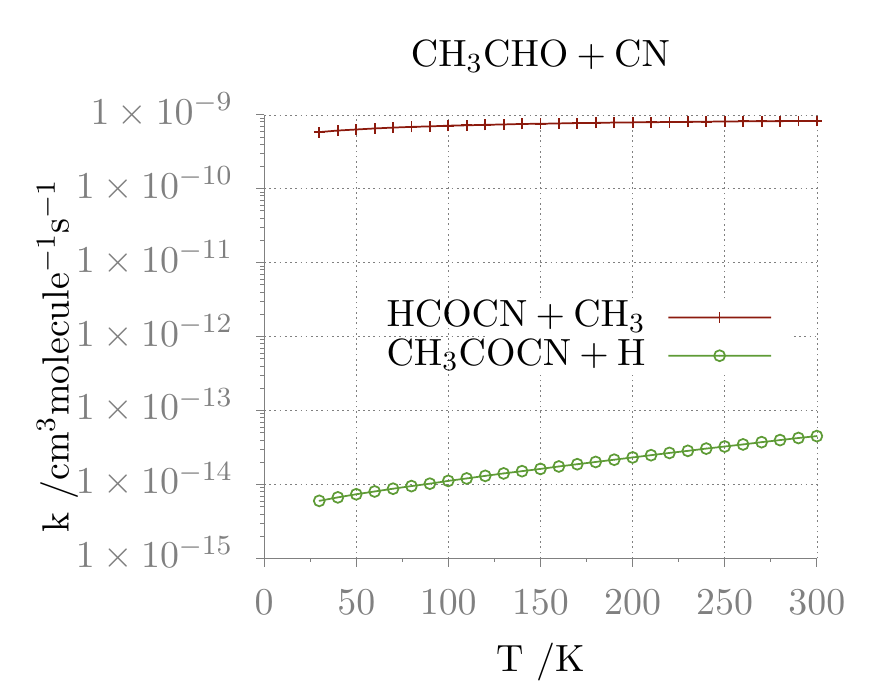}
 \end{center}
 \caption{Rate constants for the \ce{CH3CHO + CN} reaction leading to \ce{HCOCN + CH3} and \ce{CH3COCN + H}.}
 \label{fig:rateace}
\end{figure}

Global and channel specific rate constants for the \ce{CH2O + CN} and \ce{CH3CHO + CN} reactions have been computed, as described in Section \ref{Kin:Mod}, using the PESs depicted in Figures \ref{fig:1} and \ref{fig:wa}. The multi-well one-dimensional master equation has been solved by exploiting the chemically significant eigenvalues (CSEs) method within the Rice-Ramsperger-Kassel-Marcus (RRKM) approximation, as detailed in \cite{miller2006master}. All rate coefficients have been computed in the 30-300 K temperature range and at pressure of \num{1d-12} atm. The temperature dependence plots are shown in Figures \ref{fig:rateform} (channel rates) and \ref{fig:rateformglobal} (global rate) for the formaldehyde + CN reaction, and in Figure \ref{fig:rateace} for the acetaldehyde + CN reaction.

For these calculations, the CCSD(T)/CBS+CV energies, corrected by anharmonic ZPE values, have been employed for reaction paths involving a non negligible transition state, while rate constants of the barrierless channels have been computed using PST (see Section \ref{Kin:Mod} and Appendix \ref{appendix:kin}). The uncertainty on reaction rates issuing from the errors in the computed CCSD(T)/CBS+CV energies is below 20\%, which is largely sufficient for the comparison with experiment. Reaction rates computed with B2PLYP energies are within a factor of 2 from their CCSD(T)/CBS+CV counterparts in the whole range of temperatures between 50 and 300 K, except for product P2 in the acetaldehyde reaction (due to a much lower activation energy of the key step). However, since this product is negligible, the agreement is quite reasonable.

Within the temperature interval considered, for the \ce{CH2O + CN} reaction, the fastest reaction channel is always the addition path 2 (see Figure~\ref{fig:1} for path labeling), though the relevance of the abstraction path is non-negligible, its rate constant differing by less than one order of magnitude with respect to that of path 2. Regarding to the \ce{CH3CHO + CN} reaction, as expected, the dominant reaction channel at all temperatures is the one leading to HCOCN. Within the uncertainty derived from the model used for the entrance channel, the formaldehyde + CN reaction seems to be faster than the acetaldehyde + CN one in the ISM harsh conditions at least in the 30-150 K temperature range. Then, in the 150-200 K interval the global rate constants are very similar, with the reaction involving \ce{CH3CHO} becoming a little bit faster at temperatures above 200 K (see Table~\ref{tab:rate}).

Even if of limited astrophysical interest (but of potential relevance in planetary atmospheres), it is worthwhile  investigating the high-pressure limit rate constants for the interconversion of the entrance channel wells. Concerning the \ce{CH2O + CN} reaction, at 300 K, the addition entrance well interconverts into the products with a rate coefficient of \SI{2.16d-10}{\cubic\centi\metre\per\molecule\per\second} and into the reactants with a rate coefficient of \SI{1.40d-6}{\cubic\centi\metre\per\molecule\per\second}, while the abstraction entrance well never reaches in an effective way the final products, going rather back to the reactants. On the other hand, in the \ce{CH3CHO + CN} reaction, the entrance well interconversion toward the products largely overwhelms that back to the reactants, with rate coefficients of 2000 and \SI{5.86d-7}{\cubic\centi\metre\per\molecule\per\second}, respectively. This leads us to conclude that, assuming that the entrance well is collisionally stabilized, formyl cyanide can be formed only by the \ce{CH3CHO + CN} reaction, while in the ISM extreme conditions it can be formed efficiently by both reactions, since the reaction proceeds by well-skipping.

\section{Conclusion and perspectives}  

The main aim of this study was the disclosure of a feasible gas-phase mechanism for the formation of formyl cyanide in the interstellar medium. The selection of possible precursors was based on relative abundances of the cyano radical, formaldehyde, and acetaldehyde in the Sagittarius B2 region, where also formyl cyanide was detected. Guided by state-of-the-art electronic and kinetic computations, we succeeded in characterizing and validating reasonable mechanisms starting from the C-attack of the CN radical on formaldehyde as well as on acetaldehyde, and/or the abstraction of a hydrogen atom of formaldehyde by the CN radical. All the elementary steps of these reactions are ruled by energy barriers submerged with respect to reactants, thus pointing out the feasibility of HCOCN production in the extreme conditions of the ISM.
The study of the reaction mechanisms starting from acetaldehyde has shown that formation of \ce{CH3COCN}, while being feasible, is less probable than that of HCOCN. This can be explained in terms of the greater strength of the C-H bond with respect to the C-C one. 

Solution of a master equation including the different reaction channels showed that the very low pressures characterizing the ISM permit to reach the final products, even in the presence of very stable intermediates (as is the case for formaldehyde), which cannot act as effective kinetic sinks. As a consequence, under these conditions, the reaction of formaldehyde is faster than that of acetaldehyde, whereas the situation is reversed at higher pressures (where collision stabilization plays a role).
Together with the intrinsic interest of the studied systems, this paper shows, in our opinion, that state-of-the-art quantum-chemical methodologies represents a very important tool for astrochemical studies when integrated with last generation kinetic models based on the ab-initio master equation paradigm.

\acknowledgments
This work has been supported by MIUR 
(Grant Number 2017A4XRCA) and by the University of Bologna (RFO funds). The SMART@SNS Laboratory (http://smart.sns.it) is acknowledged for providing high-performance computing facilities. Support by the Italian Space Agency (ASI; `Life in Space' project, N. 2019-3-U.0) is also acknowledged.

%





\appendix

\section{Reactive PES: computational details}

(i) Preliminary investigation: Stationary points (minima and transition states) have been computed using the hybrid B3LYP functional \citep{{Becke1993},{Lee1988}} in conjunction with SNSD basis set \citep{Barone2008}, and accounting for dispersion effects according to Grimme's D3 scheme \citep{Grimme2010} combined with Becke-Johnson (BJ) damping function \citep{Grimme2011}. The nature of all stationary points found on the PES has been checked by diagonalizing the corresponding Hessian matrices. To correctly connect two minima through the corresponding transition state, intrinsic reaction coordinate (IRC) calculations have been performed \citep{Fukui1981,Hratchian2011}.

(ii) Definition of the reactive PES: All stationary points have been re-investigated at the B2PLYP/maug-cc-pVTZ-\emph{d}H level \citep{Fornaro2016,papajak2009,doi:grimme2006}, always employing the D3BJ correction. At this stage, energies are corrected for the ZPE contribution within the harmonic approximation.

(iii) Final thermochemistry: The most energetically favored paths have been selected. For the corresponding stationary points, single-point energy calculations, at the B2PLYP/maug-cc-pVTZ-\emph{d}H reference geometries, have been performed by means of the CCSD(T)/CBS+CV composite scheme \citep{Miriam2,hosop,Barone2013}, which is based on CC theory employing the singles and doubles approximation (CCSD) augmented by a perturbative treatment of triple excitations, CCSD(T) \citep{{Raghavachari2013},{Watts1993}}. 

The CCSD(T)/CBS+CV energy is obtained according to the following expression:
\begin{equation}\label{composito}
E_{tot}=E^{\infty}_{HF} + \Delta E^{\infty}_{CCSD(T)}+ \Delta E_{CV},
\end{equation}
where the first term on the right-hand side is the Hartree-Fock self-consistent-field (HF-SCF) energy extrapolated to the complete basis set (CBS) limit ($E^{\infty}_{HF}$) by means of the exponential extrapolation formula by Feller \citep{Feller1993}.
The second term is the extrapolation to the CBS limit of the CCSD(T) correlation energy ($\Delta E^{\infty}_{CCSD(T)}$), using the 2-point \emph{n}$^{-3}$ formula by \cite{Helgaker1997}. In order to exploit these extrapolative expressions, HF-SCF and CCSD(T) energy calculations have been performed using Dunning's cc-pV$n$Z basis sets \citep{Dunning-JCP1989_cc-pVxZ}, with $n$=T,Q,5 for the HF-SCF extrapolation and $n$=T,Q for the CCSD(T) correlation energy. The last term is the core-valence (CV) correlation energy correction ($\Delta E_{CV}$), which is evaluated as the energy difference between all-electron and frozen-core CCSD(T) calculations in the same basis set (cc-pCVTZ; \cite{WD95}). This contribution is required because the extrapolation to the CBS limit is performed within the frozen-core approximation.
As mentioned in the main text, the CCSD(T)/CBS+CV energies have been combined with anharmonic B2PLYP ZPE corrections obtained within second-order vibrational perturbation theory (VPT2; \cite{mills1972molecular,Hoy1972,Barone2004}).
Among the various formulations, the Hybrid Degeneracy Corrected VPT2 (HDCPT2) model \citep{Bloino2012} has been used.

\section{Kinetic calculations: computational details}\label{appendix:kin}

As aforementioned in Section \ref{Kin:Mod}, barrierless channels were treated with PST. Indeed, it provides a useful, and easily implemented, reference theory for barrierless reactions. The basic assumption in PST is that the interaction between the two reacting fragments is isotropic and does not affect the internal fragment motions. This assumption is only valid if the dynamical bottleneck lies at large separations where the interacting fragments have free rotations and unperturbed vibrations, which is generally true for low temperature phenomena as those occurring in the ISM. The isotropic potential is assumed to be described by the functional form $-\frac{C}{R^6}$, where the coefficient $C$ is obtained by fitting the energies obtained at various long-range distances of fragments. The latter were obtained using the double-hybrid B2PLYP functional in conjunction with the maug-cc-pVTZ-\emph{d}H basis set and incorporating dispersion corrections (D3BJ as already mentioned above). The $C$ coefficients were then corrected in order to obtain entrance channels rate constant values scaled by a 0.9 factor (in analogy to what done in  \cite{doi:10.1063/1.1539035}), which is a dynamical correction factor that takes into account the recrossing effects. The final values for the $C$ are 290.79, 443.34 and \SI{506.68}{\bohr\tothe{6}\hartree}, for the formaldehyde addition, formaldehyde abstraction and acetaldehyde, respectively. To check the role of the basis set superposition error in the evaluation of the PST coefficient, counterpoise (CP) corrected long-range electronic energies have been computed at various interfragment distances, thereby leading to a small difference of ca. \SI{3}{\bohr\tothe{6}\hartree} with respect to the non-CP corrected values. Such a small difference is essentially non-relevant for the calculation of the entrance reactive flux, thus allowing to safely use non-CP corrected energies.

In Table \ref{tab:rate}, the channel specific and global rate constants are collected.

\begin{table}[ht!]
\centering
\caption{Product-formation rate constants (in cm$^3$ molecule$^{-1}$ s$^{-1}$) at \num{1d-12} atm as a function of temperature.}
\label{tab:rate}
\resizebox{\textwidth}{!}{%
\begin{tabular}{@{}ccccccc|cc@{}}
\toprule
\multicolumn{1}{c}{} & \multicolumn{6}{c}{Formaldehyde} & \multicolumn{2}{|c}{Acetaldehyde} \\
\toprule
T {[}K{]} & Path 1   & Path 2   & Path 3   & Path 4   & Abstraction & Global & \ce{HCOCN + CH3}       & \ce{CH3COCN + H}   \\ \midrule
30        & \num{7.42d-10} & \num{1.03d-09} & \num{8.04d-10} & \num{3.42d-11} & \num{1.46d-10}    & \num{8.35d-10} & \num{5.81d-10} & \num{6.03d-15}\\
40        & \num{8.70d-10} & \num{1.05d-09} & \num{6.99d-10} & \num{1.68d-11} & \num{1.00d-10}    & \num{1.02d-09} & \num{6.10d-10} & \num{6.70d-15}\\
50        & \num{8.86d-10} & \num{1.06d-09} & \num{5.97d-10} & \num{9.63d-12} & \num{7.64d-11}    & \num{1.15d-09} & \num{6.33d-10} & \num{7.37d-15}\\
60        & \num{8.23d-10} & \num{1.05d-09} & \num{5.06d-10} & \num{6.17d-12} & \num{6.27d-11}    & \num{1.13d-09} & \num{6.52d-10} & \num{8.06d-15}\\
70        & \num{7.58d-10} & \num{1.03d-09} & \num{4.30d-10} & \num{4.21d-12} & \num{5.44d-11}    & \num{1.11d-09} & \num{6.69d-10} & \num{8.78d-15}\\
80        & \num{6.93d-10} & \num{1.00d-09} & \num{3.67d-10} & \num{3.02d-12} & \num{4.93d-11}    & \num{1.08d-09} & \num{6.84d-10} & \num{9.54d-15}\\
90        & \num{6.33d-10} & \num{9.70d-10} & \num{3.15d-10} & \num{2.25d-12} & \num{4.60d-11}    & \num{1.05d-09} & \num{6.97d-10} & \num{1.03d-14}\\
100       & \num{5.78d-10} & \num{9.34d-10} & \num{2.72d-10} & \num{1.73d-12} & \num{4.41d-11}    & \num{1.02d-09} & \num{7.10d-10} & \num{1.12d-14}\\
110       & \num{5.27d-10} & \num{8.97d-10} & \num{2.37d-10} & \num{1.36d-12} & \num{4.31d-11}    & \num{9.88d-10} & \num{7.21d-10} & \num{1.21d-14}\\
120       & \num{4.82d-10} & \num{8.60d-10} & \num{2.08d-10} & \num{1.10d-12} & \num{4.28d-11}    & \num{9.54d-10} & \num{7.31d-10} & \num{1.31d-14}\\
130       & \num{4.42d-10} & \num{8.22d-10} & \num{1.83d-10} & \num{9.01d-13} & \num{4.29d-11}    & \num{9.19d-10} & \num{7.40d-10} & \num{1.41d-14}\\
140       & \num{4.05d-10} & \num{7.86d-10} & \num{1.62d-10} & \num{7.54d-13} & \num{4.35d-11}    & \num{8.86d-10} & \num{7.49d-10} & \num{1.52d-14}\\
150       & \num{3.73d-10} & \num{7.51d-10} & \num{1.45d-10} & \num{6.34d-13} & \num{4.44d-11}    & \num{8.54d-10} & \num{7.57d-10} & \num{1.63d-14}\\
160       & \num{3.44d-10} & \num{7.17d-10} & \num{1.30d-10} & \num{5.46d-13} & \num{4.56d-11}    & \num{8.23d-10} & \num{7.64d-10} & \num{1.75d-14}\\
170       & \num{3.18d-10} & \num{6.85d-10} & \num{1.16d-10} & \num{4.65d-13} & \num{4.70d-11}    & \num{7.93d-10} & \num{7.71d-10} & \num{1.88d-14}\\
180       & \num{2.95d-10} & \num{6.54d-10} & \num{1.06d-10} & \num{4.10d-13} & \num{4.86d-11}    & \num{7.65d-10} & \num{7.77d-10} & \num{2.02d-14}\\
190       & \num{2.74d-10} & \num{6.26d-10} & \num{9.63d-11} & \num{3.64d-13} & \num{5.04d-11}    & \num{7.38d-10} & \num{7.83d-10} & \num{2.17d-14}\\
200       & \num{2.55d-10} & \num{5.98d-10} & \num{8.80d-11} & \num{3.21d-13} & \num{5.23d-11}    & \num{7.13d-10} & \num{7.89d-10} & \num{2.33d-14}\\
210       & \num{2.38d-10} & \num{5.73d-10} & \num{8.06d-11} & \num{2.90d-13} & \num{5.45d-11}    & \num{6.89d-10} & \num{7.94d-10} & \num{2.49d-14}\\
220       & \num{2.22d-10} & \num{5.48d-10} & \num{7.42d-11} & \num{2.62d-13} & \num{5.67d-11}    & \num{6.67d-10} & \num{7.99d-10} & \num{2.67d-14}\\
230       & \num{2.08d-10} & \num{5.26d-10} & \num{6.84d-11} & \num{2.33d-13} & \num{5.91d-11}    & \num{6.46d-10} & \num{8.03d-10} & \num{2.86d-14}\\
240       & \num{1.96d-10} & \num{5.04d-10} & \num{6.33d-11} & \num{2.15d-13} & \num{6.16d-11}    & \num{6.27d-10} & \num{8.07d-10} & \num{3.06d-14}\\
250       & \num{1.84d-10} & \num{4.84d-10} & \num{5.88d-11} & \num{1.96d-13} & \num{6.43d-11}    & \num{6.09d-10} & \num{8.11d-10} & \num{3.27d-14}\\
260       & \num{1.73d-10} & \num{4.65d-10} & \num{5.48d-11} & \num{1.81d-13} & \num{6.71d-11}    & \num{5.92d-10} & \num{8.14d-10} & \num{3.49d-14}\\
270       & \num{1.64d-10} & \num{4.47d-10} & \num{5.11d-11} & \num{1.67d-13} & \num{6.99d-11}    & \num{5.76d-10} & \num{8.17d-10} & \num{3.73d-14}\\
280       & \num{1.55d-10} & \num{4.30d-10} & \num{4.78d-11} & \num{1.58d-13} & \num{7.29d-11}    & \num{5.61d-10} & \num{8.20d-10} & \num{3.98d-14}\\
290       & \num{1.46d-10} & \num{4.14d-10} & \num{4.49d-11} & \num{1.47d-13} & \num{7.60d-11}    & \num{5.47d-10} & \num{8.22d-10} & \num{4.25d-14}\\
300       & \num{1.39d-10} & \num{3.98d-10} & \num{4.22d-11} & \num{1.36d-13} & \num{7.92d-11}    & \num{5.35d-10} & \num{8.24d-10} & \num{4.53d-14}\\ \bottomrule
\end{tabular}
}
\end{table}

\section{Software packages}
All DFT and VPT2 calculations have been carried out with the Gaussian software \citep{g16}, while for those based on CCSD(T) the CFOUR program \citep{cfour,cfour-new} has been employed. RRKM simulations have been performed with MESS code \citep{georgievskii2013reformulation}, available at https://github.com/PACChem/MESS.

\bibliography{Astrochem}{}
\bibliographystyle{aasjournal}



\end{document}